\documentclass[%
 reprint,
 amsmath,amssymb,
 aps,
floatfix,superscriptaddress
]{revtex4-2}

\usepackage{graphicx}
\usepackage{dcolumn}
\usepackage{bm}
\usepackage{xcolor}
\usepackage{soul} 

\begin{document}

\preprint{APS/123-QED}

\title{Observation of the p-wave Shape Resonance}
\author{Baruch Margulis}
\affiliation{Department of Chemical and Biological Physics, Weizmann Institute of Science, Rehovot, Israel}
\author{Prerna Paliwal}
\affiliation{Department of Chemical and Biological Physics, Weizmann Institute of Science, Rehovot, Israel}
\author{Wojciech Skomorowski}
\affiliation{Centre of New Technologies, University of Warsaw, Banacha 2c, 02-097 Warsaw, Poland}
\author{Mariusz Pawlak}
\affiliation{Faculty of Chemistry, Nicolaus Copernicus University in Toru\'n, Gagarina 7, 87-100 Toru\'n, Poland}
\author{Piotr S. \.Zuchowski}
\affiliation{Institute of Physics, Faculty of Physics, Astronomy and Informatics, Nicolaus Copernicus University, Grudzi\c adzka 5, 87-100, Toru\'n, Poland}
\author{Edvardas Narevicius}
 \email{email: edvardas.narevicius@weizmann.ac.il}
\affiliation{Department of Chemical and Biological Physics, Weizmann Institute of Science, Rehovot, Israel}

\date{\today}

\begin{abstract}
{We observe a $p$-wave resonance in collisions between metastable neon and rovibrationally ground state HD molecules at a collision energy of $k_B\times$22~mK. This is the first observation of the lowest quasibound angular momentum state in molecular collisions. This measurement was enabled by the reduction of the lowest collision energy achieved without laser cooling, using the phase-space correlation in cold molecular beams. We demonstrate that contrary to higher $l$-state resonances, the $p$-wave resonance allows sensitive probing of the leading term of the van der Waals intermolecular interaction. Furthermore, the same sensitivity to the long range part of the interaction potential allows an accurate estimation of $p$-wave resonance lifetime using only fundamental constants and the dispersion coefficient.}\end{abstract}

\maketitle

Many processes in nature at low collision energies are slow and occur mainly via quantum resonance states. For example, collision products can be either blocked by a~potential barrier as in F+H$_2$ reaction \cite{kim2015spectroscopic,tizniti2014rate} or be suppressed by quantum symmetry as in the case of atom-ion spin exchange \cite{sikorsky2018phase}, molecular fine structure changing collisions~\cite{chefdeville2013observation} and Penning ionization (PI) reactions \cite{lavert2014observation}. Shape type resonances in particular are identified by a~well-defined angular momentum quantum number and are formed by tunneling through a centrifugal barrier which arises due to the addition of the centrifugal term $\frac{\hbar^2l(l+1)}{2\mu R^2 }$ to the intermolecular potential. Partial wave resonances in molecular scattering experiments were first observed in elastic scattering \cite{toennies1979molecular} and later in reactive and inelastic collisions using merged beams technique \cite{henson2012observation} as well as in low intersection angle collisions \cite{chefdeville2013observation}. These resonances are excellent probes of intermolecular potentials~\cite{vogels2018scattering}, enable the direct probe of interaction anisotropy~\cite{klein2017directly}, and serve as a benchmark for modern quantum chemistry theories~\cite{csaszar2020rotational}. 
Observation of partial wave resonances require collision energies of $k_B\times1$~K and below. The low collision energies limit the number of accessible angular momentum states, enabling observation of a well resolved resonance state, eventually reaching the point where collision complex possesses a single quanta of angular momentum. This allows a full control not only over the internal molecular degrees of freedom but also the intermolecular ones, constraining all the quantum numbers involved in the collision. To date, experiments using small angle scattering were able to probe the onset of a $p$-wave resonance~\cite{de2020imaging}.

Here we present the first measurement of the lowest possible angular momentum quasibound state, the $p$-wave resonance state, in molecular scattering. We demonstrate that such low angular momentum states are particularly important in sensing the leading terms in the van der Waals part of the long range intermolecular interaction. The same sensitivity allows us to accurately estimate the $p$-wave resonance lifetime without any knowledge of short range physics using a semiclassical approach together with a truncated potential. This approach is complementary to the estimation of the low energy scattering length \cite{gribakin1993calculation}. This is a useful result, since in many experiments the measured resonance linewidth  does not necessarily correspond to lifetime, it may be limited by the energy resolution. Using our method we can now estimate the natural $p$-wave resonance lifetime from fundamental constants and the dispersion parameter alone. 

We detect the fully resolved $p$-wave resonance in PI collisions between metastable neon atoms and neutral hydrogen deuteride molecules at a collision energy of $k_B\times(22.2\pm 5.4)$~mK. We observe this resonance state using the merged beam technique, taking advantage of the phase space correlation within two overlapping particle bunches with nearly identical mean velocities.

\begin{figure*}[t]
\centering
\includegraphics[width=1\textwidth]{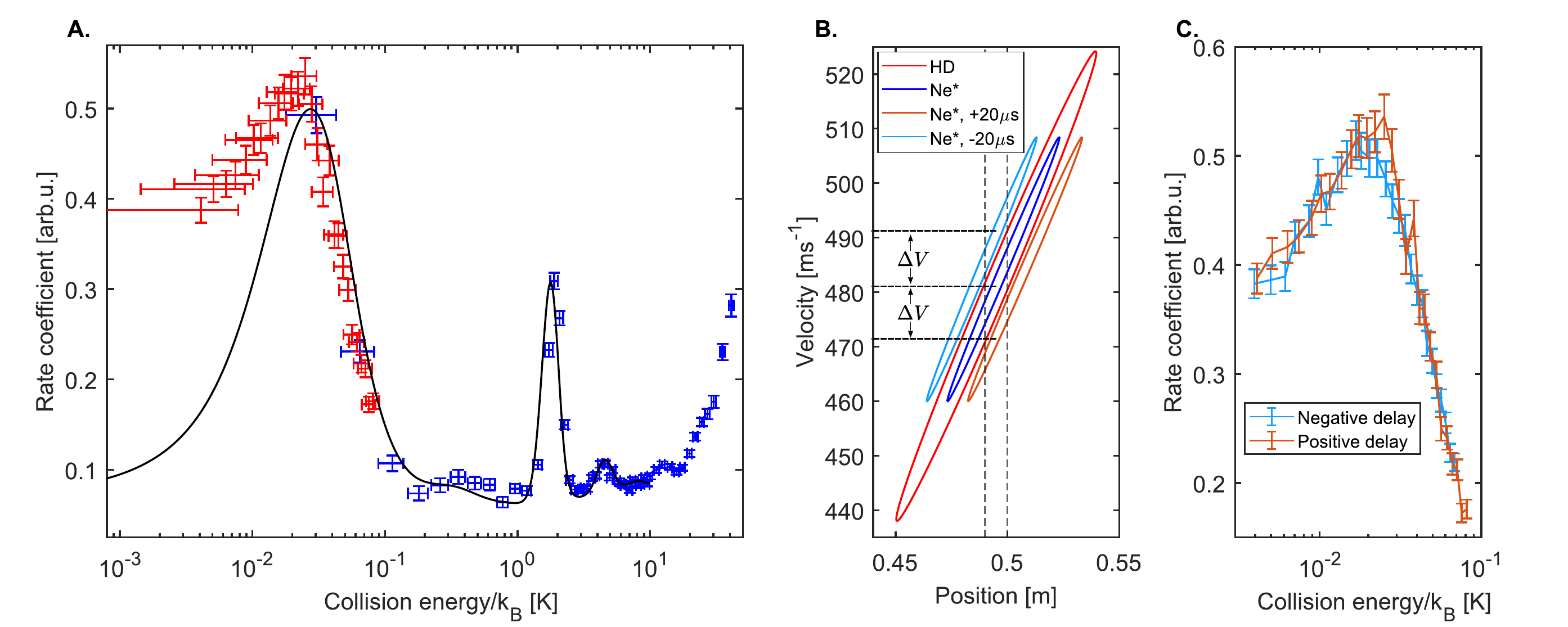} 
\caption{A. The data points with error bars represent measured rate coefficients for the Ne$^*$--HD PI reaction and their uncertainties. The black curve depicts the calculated rate coefficients using the interaction potential, obtained at the CCSD(T) level of theory with additional 2.9\% scaling of the electron correlation energy. Vertical error bars represent statistical errors. Horizontal error bars represent the uncertainty in collision energy as derived from the uncertainty in individual beam velocity. B. Phase space distributions snapshot of the Ne$^*$ and HD beams assuming the experimental velocity distributions at the low-energy regime in the interaction region of our detector. For the red curve, the delay between the beams is set for a maximal overlap at the center of the VMI detector. Blue and orange contours represent a shift of $\pm$20~$\mu$s in delay between the valves, respectively. Vertical dotted lines represent the edges of the detection region. C. Comparison of the measured shape of the low-energy resonance for negative and positive beam delay scans.} 
\label{fig1}
\end{figure*}

The experimental setup is described in detail elsewhere~\cite{paliwal2021determining}. In short, two pulsed supersonic beams of pure Ne and a 50/50 mixture of HD and Ne are generated by two Even-Lavie valves~\cite{even2000cooling}. The Ne beam is excited to the $^3P_2$ metastable state (Ne$^*$) by a dielectric barrier discharge which is located directly after the valve nozzle~\cite{luria2009dielectric}. The Ne$^*$ beam is merged with the HD beam using a curved 20 cm magnetic guide leading to a zero relative angle between the beams. Product ions are extracted and detected using a Velocity Map Imaging (VMI) spectrometer, which operates in ion count mode. HD molecules are generated in their ground rovibrational state by supersonic expansion. The Ne$^*$ beam velocity and standard deviation, $\sigma$, are characterized by measuring the time-of-flight (TOF) information using an on-axis multichannel plate. The HD beam is characterized using a TOF mass spectrometer (TOF-MS) which contains an electron impact ionization element. The collision energy is determined by the mean relative velocity between the beams. The rate coefficient is calculated according to the ratio between counts of HD$^+$ ions and the intensity of the reactant beams at the intersection point. Every measurement of product ions is followed by a measurement of background ions, which is deducted from the overall count. 

We present the measured rate coefficient as a function of collision energy together with the results of theoretical calculations in figure~\ref{fig1}. The experimental data consists of two sets of measurements. For the first set, the collision energy is tuned by manipulating the mean value of the velocity of the HD beam while simultaneously setting the TOF delay between the valves such that the beams merge at the center of the detection region. The first set is represented by blue data points and includes collision energies between $k_B\times$30~mK and $k_B\times$40~K. For this set, the Ne$^*$ beam velocity is kept constant at 470~m/s ($\sigma=11.2$~m/s) while the HD beam velocity is tuned between 485~m/s ($\sigma=35$~m/s) and 982~m/s ($\sigma=49$~m/s) by changing the HD valve temperature between 64 and 210~K. Each data point represents events accumulated over 5000 repetitions. 
Our merged beam setup allows us to scan collision energy in a low energy range with high resolution using the phase space correlation, which develops during the beam propagation from the source to the detection region~\cite{shagam2013sub}. Since beam generation time is much shorter as compared to the time of flight, velocity distribution in the detection volume becomes squeezed, with the standard deviation of 2~m/s for the Ne$^*$ beam. In order to scan the relative velocity between two such beams, we can either change the center velocity of one beam relative to the other as is done in most experiments or to change the time delay between two near-matched velocity beams as shown in figure~\ref{fig1}B. The second data set acquired using phase space correlation is represented by red data points for collision energies between $k_B\times$4~mK and $k_B\times$75~mK. For this set, both the Ne* and HD beams' velocities are kept constant at 484~m/s ($\sigma=11$~m/s) and 475~m/s ($\sigma=33$~m/s), respectively. The collision energy is tuned by changing the time delay between the valves opening time in the range of 0 to 45~$\mu$s. Each point represents data acquired in 10000 repetitions.
By shifting the valve opening time of one beam relative to the other, once in the positive direction and once in the negative direction, we are able to probe the low energy regime in two separate measurements (figure \ref{fig1}C). This allows us to obtain an accurate value for the HD beams' generation time therefore reducing the uncertainty in the TOF of the HD beam from around 20~$\mu$s, which is the valve opening time, to 0.5~$\mu$s. For the Ne$^*$ beam, the generation time is accurately defined by the ignition instance of the dielectric discharge which is 45~$\mu$s. 

\begin{figure}[t]
\centering
\includegraphics[width=0.49\textwidth]{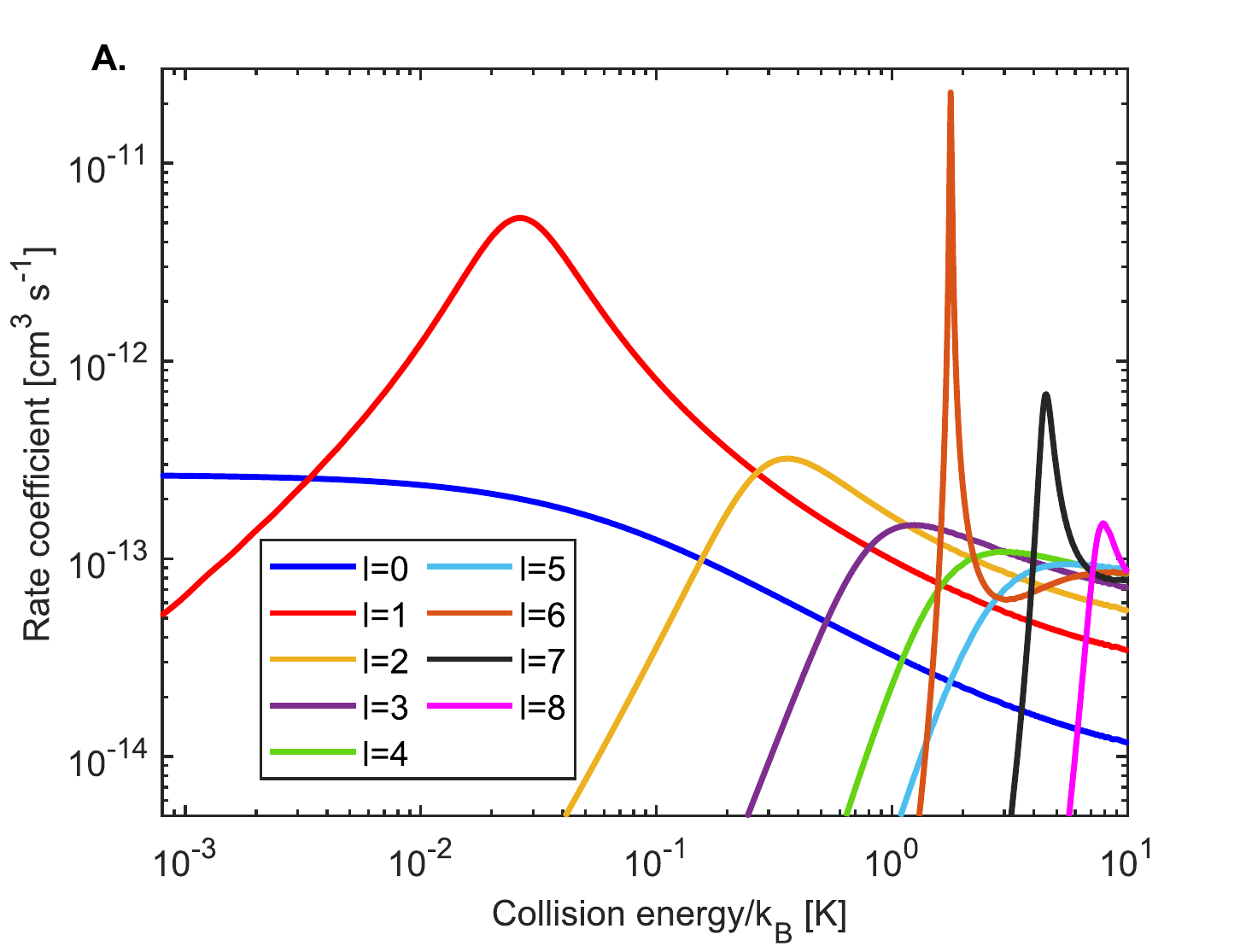} 
\includegraphics[width=0.49\textwidth]{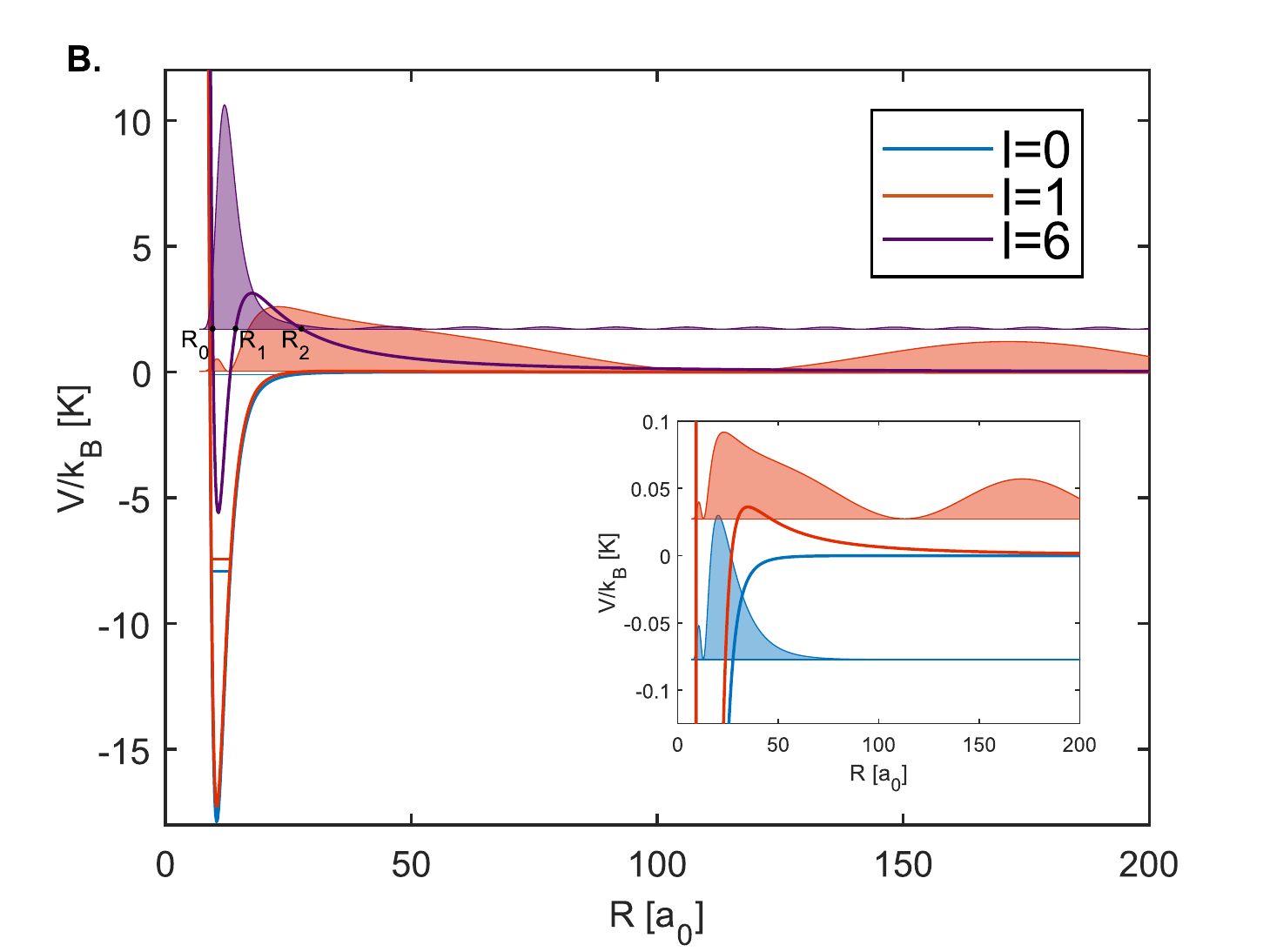} 
\caption{A. Partial wave selected PI rate coefficient as a function of collision energy. B. Effective potentials for the $l=0,1$ and $6$ states of collision. Resonance wave functions for the partial waves $l=1$ and $l=6$ are shown by shaded orange and purple areas, respectively. Bound state energies for $l=0$ and $l=1$ are depicted by orange and blue horizontal lines. The inset shows a zoomed-in view where the $p$-wave barrier can be seen. The wave function of the excited bound state for $l=0$ is presented as a shaded blue curve. Wave functions and energies are calculated using discrete variable representation~\cite{colbert1992novel}.} 
\label{fig2}
\end{figure}

Theoretical rate coefficients are estimated using an {\it ab initio} complex-valued potential energy surface. The real part of the potential is calculated using the supermolecular approach and the coupled-cluster method with single, double, and approximate non-iterative triple excitations [CCSD(T)] in a very large, diffused basis set (doubly augmented correlation consistent polarized valence quintuple zeta with midbond functions). The imaginary part of the potential is obtained using the Fano-Feshbach approach combined with equation-of-motion coupled-cluster single and double wave function (EOM-CCSD)~\cite{skomorowski21}, and with a bound-domain projection operator adapted to the PI problem. To construct final (continuum) electronic states of the system in the Feshbach theory, the orbital describing the outgoing electron is approximated with a Coulomb wave centered at the midpoint of the HD molecule. The primary source of uncertainty in the real part of the interaction potential is the electron correlation energy which depends on the truncation of the basis set and the coupled-cluster excitation accuracy level. The electron correlation energy is adjusted to match the experimental findings by scaling it by 1.029. The rotational wave function of the HD molecule in the ground rotational state is spherically symmetric, and the anisotropy resulting from the spatial distribution of the excited metastable neon atom is negligible, similar to the interactions with He and Ar atoms~\cite{hapka2013}. Hence, only the isotropic part of the intermolecular potential, expanded in renormalized spherical harmonics, is considered in the calculation. To this end, the {\em ab initio} calculations are carried out using the linear and T-shaped geometries of the complex. The scattering wave function is propagated from 7 to 500~$a_0$, where $a_0$ is the Bohr radius, applying the log-derivative method~\cite{Johnson_1973}.
The reaction rate coefficient is convoluted according to the experimental resolution which is expressed by the spread in relative velocity per collision energy. The latter is estimated using 1D particle trajectory simulations, taking into account the experimental spread in individual beam velocities and the size of the VMI extraction area. The theoretical result is shown as a black curve in figure~\ref{fig1}A. 

We observe three scattering resonances at $k_B\times(22.2\pm5.4)$~mK, $k_B\times (1.92\pm0.13)$~K, and $k_B\times (4.51 \pm 0.20)$~K. Figure~\ref{fig2} presents the calculated partial wave contribution to each resonance state and the numerically calculated resonance state wave functions. We identify the resonance at $k_B\times$22~mK as arising from the $p$-wave partial wave state. The higher energy resonances at $k_B\times$1.9~K and $k_B\times$4.5~K arise from the $l=6$ and $l=7$ states accordingly.

Surprisingly, the $p$-wave and $l=6$ resonances have nearly identical widths although the $l=6$ resonance is stronger confined by the centrifugal barrier.
Shape type resonance width, $\Gamma$, which is inversely proportional to the lifetime $\tau=\hbar/\Gamma$, can be estimated using semiclassical Gamow approximation, which introduces two parameters. One is the tunneling probability through a centrifugal barrier whereas the second parameter defines the timescale given by the nuclear motion frequency~\cite{gamow1928quantentheorie,gurvitz1987decay}:
\begin{equation}
\Gamma = \frac{\hbar^2}{2\mu} \left [ \int_{R_0}^{R_1}\frac{1}{k(R)}dR \right ]^{-1} \exp{\left[-2\int_{R_1}^{R_2}|k(R)|dR\right]},  
\end{equation}
where $R_0$, $R_1$, and $R_2$ are the classical turning points as indicated for the $l=6$ resonance state in figure~\ref{fig2}B.  $k(R)=\sqrt{2\mu(E-V(R))}/\hbar$ is the semiclassical wave vector. 

For resonance states where tunneling probability is $\ll$1, this leads to a characteristic exponential decay or a Lorentzian lineshape in the energy spectrum. A very unusual situation arises for resonances that are located near the top of the centrifugal barrier where tunneling probability approaches unity. In such a~case the decay time is mainly defined by the time it takes particles to reach the centrifugal barrier. For $p$-wave resonances, the top of the centrifugal barrier is located near the threshold. This leads to both high tunneling probability and slow nuclear dynamics since particles spend most of the time exploring the long range part of the interaction potential. As a result, the linewidth of such shape resonances can be orders of magnitude narrower as compared to orbiting resonances that have been detected earlier \cite{paliwal2021determining}.

\begin{table}[t]
\caption{\label{Gamma_tab} 
Resonance positions and widths calculated by the CAP method ($E_{\rm res}=E_{\rm exact}-i\frac{\Gamma_{\rm exact}}{2}$), semiclassical widths obtained using exact resonance energies ($\Gamma_{\rm sc}$), tunneling probabilities ($P_{\rm sc}$), and widths for the truncated potential ($\Gamma_{\rm sc}^{\rm trunc}$).
All energy values are in units of Kelvin.}
\begin{ruledtabular}
\begin{tabular}{cccccc}
$l$ & $E_{\rm exact}$ & $\Gamma_{\rm exact}$ & $ P_{\rm sc}$ & $\Gamma_{ \rm sc}$ & $\Gamma_{\rm sc}^{\rm trunc}$\\
\hline
1 & 0.023    & 0.030   & 0.50 \, & 0.045   & 0.046   \\
6 & 1.781    & 0.033   & 0.023   & 0.036   & 0.047   \\
7 & 4.47 \,  & 0.59 \, & 0.59 \, & 0.64 \, & 0.83 \, \\
\end{tabular}
\end{ruledtabular}
\end{table}

A comparison of the numerically exact width as calculated using complex absorbing potential \cite{riss1993calculation,santra2002non} and semiclassical approximation is given in table \ref{Gamma_tab}. The semiclassical approximation reliably estimates the width within an order of magnitude.
The $p$-wave and $l=6$ resonance lifetimes are nearly identical despite the order of magnitude difference in tunnelling probabilities. Since the tunneling probability for the $p$-wave resonance approaches unity the lifetime becomes limited by the timescale of the nuclear motion behind the barrier, given by the first term in (1). For the $p$-wave resonance the slow nuclear dynamics timescale results in a resonance width which is larger as compared to the resonance energy position leading to a distinct non-lorentzian form of the resonance peak indicating a non-exponential decay. For the $l=7$ resonance state, which has a similar magnitude of tunneling probability, the real part of the resonance energy is by two orders of magnitude higher leading to a 20 fold shorter lifetime. 

\begin{figure}[t]
\centering
\includegraphics[width=0.49\textwidth]{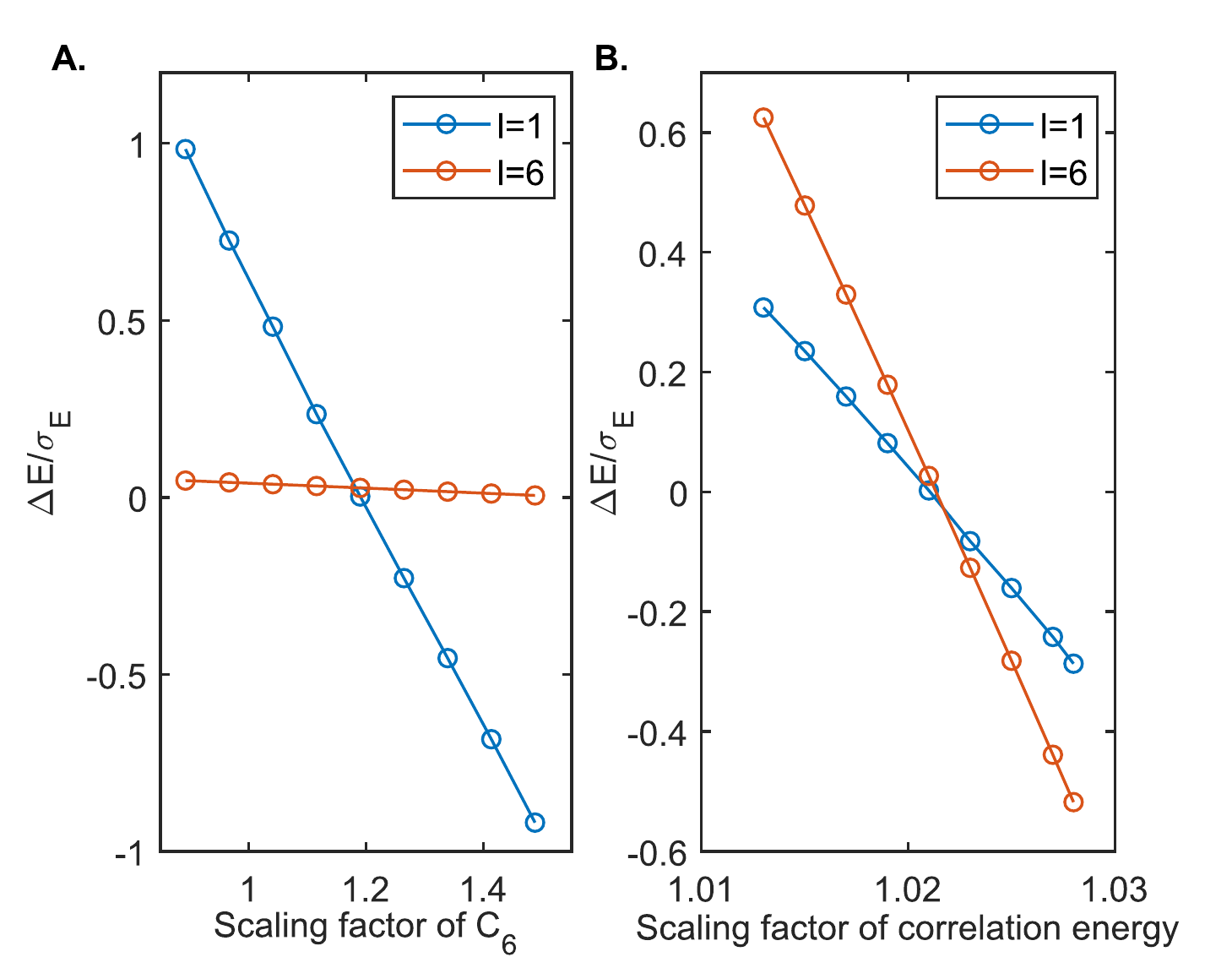} 
\includegraphics[width=0.49\textwidth]{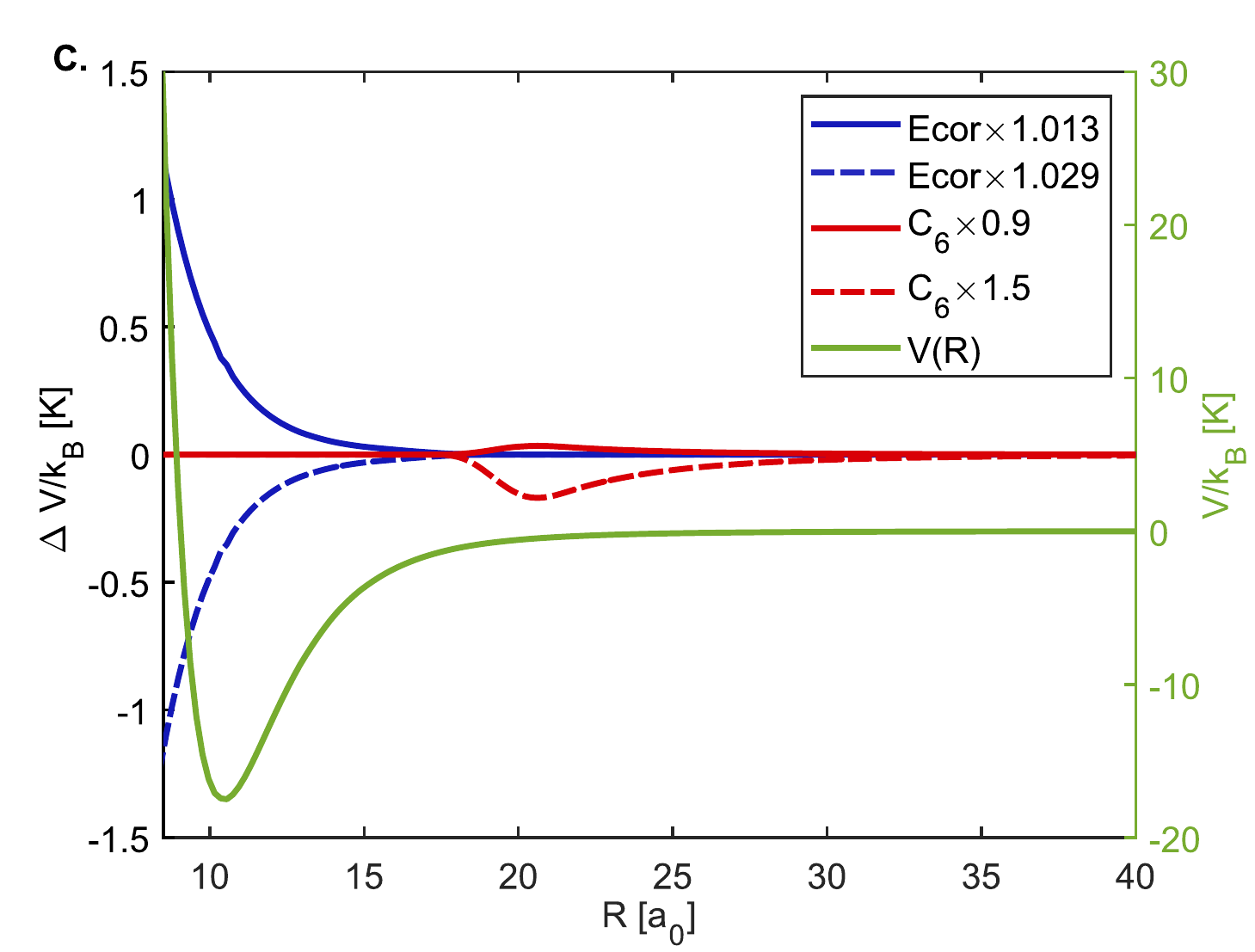}
\caption{Shift in position of $l=1$ and $l=6$ resonances as a function of scaling  the $C_6$ coefficient (A) and  electron correlation energy (B). The shift is expressed in the units of the width of the measured resonance ($\sigma_E$). For $C_6$ scaling, the potential is smoothly modified between $R=17$ and $R=22$~$a_0$. Calculations are performed around scaling values which provide the lowest error for both resonance positions simultaneously. C. Addition to the intermolecular potential due to scaling of the electron correlation energy (blue curves) and scaling of $C_6$ (red curves). The interaction potential is presented by a green curve.
}
\label{fig3}
\end{figure}

The $p$-wave centrifugal barrier is located at intermolecular distance that is described by the long-range dispersion interaction alone. Since the $p$-wave resonances are located close to the threshold we show that lifetime can be calculated using the semiclassical approximation with a truncated potential which includes only the long range part and the centrifugal term: $V_{\rm lr}(R) =-\frac{C_6}{R^6}-\frac{C_8}{R^8}+\frac{\hbar^2 l(l+1)}{2\mu R^2}$. We choose the truncation position at a potential minimum, although our result is not sensitive to the exact truncation point. For the $p$-wave resonance, the semiclassical lifetime evaluated using only the long range part of the interaction agrees well with the full calculation, resulting in a relative error of less than 1\%. This demonstrates that the lifetime of $p$-wave resonance is nearly universal quantity that can be estimated using the reduced mass and the leading terms in the long range interaction potential.

Whereas $p$-wave resonance lifetime is sensitive only to the long range part of the interaction, the energy position depends both on the short and the long range parts of the interaction. However, since the two lowest energy resonance states occupy different areas in the phase space we expect different sensitivity to the interaction potential details.
An accurate description of shape resonance position requires potential energy surface exceeding the accuracy achievable with the ``gold standard'' of current quantum chemistry methods \cite{csaszar2020rotational}. As a result, calculated potential surfaces are usually uniformly scaled within the estimated error limits \cite{wallis2011prospects,chefdeville2012appearance} or modified only at the short range by scaling the electron correlation energy  \cite{klein2017directly}. By independently scaling the electron correlation energy and the leading van der Waals coefficient ($C_6$), we separately modify the short and long range parts of the interaction potential, respectively, as shown in  figure~\ref{fig3}. The latter allows us to demonstrate that different angular momentum shape resonances probe different parts of the interaction potential.

For every modification of the interaction potential, we calculate the position of the $l=6$ and $l=1$ resonances and express the shift in resonance position in units of the measured resonance width ($\sigma_E$) which is limited by the experimental resolution. The results are shown in figures~\ref{fig3}A and \ref{fig3}B. Scaling of the correlation energy in the range from 1.013 to 1.029 resulted in a shift of $1.1\sigma_E$ for the $l=6$ state and $0.6\sigma_E$ for the $l=1$ state (figure~\ref{fig3}B). On the other hand, scaling of the $C_6$ coefficient by 0.9 to 1.5 has altered the position of the $p$-wave resonance by 1.9$\sigma_E$, whereas the $l=6$ resonance was shifted only by $0.04\sigma_E$ (figure~\ref{fig3}A). While the position of the $l=6$ resonance is by a factor of 2 more sensitive to the short range interaction, the position of the $p$-wave resonance is by a factor of 50 more sensitive to the strength of the long range interaction. The simultaneous probing of the shape resonance phenomenon in the few kelvin and millikelvin energy regimes effectively probes the intermolecular potential at all regions of the interaction.

The experimental ability to constrain a collision process to a single, low-value partial wave state will further improve the imaging resolution in molecular collision experiments. The narrow distribution of initial quantum states may allow to unravel a more complex, post-ionization dynamics of the ion-neutral system enabling, for example, the detection of different Feshbach resonances which are located close to the dissociation threshold~\cite{margulis2020direct}. This requirement is essential for applying the coincidence detection technique for molecular PI collisions, where information about the spectrum of Feshbach resonances and their multi-channel decay probabilities is blurred by the increasing number of accessible angular momentum states.

\section*{Acknowledgments}
We thank Christiane Koch and Yuval Shagam for helpful discussions. B.M., P.P. and E.N. acknowledge funding from the Israel Science Foundation. P.S.Z. is grateful to National Science Centre of Poland for funding the project Sonata Bis 9 (2019/34/E/ST4/00407). The work of W.S. was supported by the Polish National Agency for Academic Exchange within Polish Returns Programme.

\newpage

\bibliography{main}

\end{document}